\documentclass[a4paper,12pt]{article}
\usepackage{graphicx}
\usepackage[top=3.0cm,bottom=3.0cm,left=3.0cm,right=2.0cm]{geometry}
\usepackage{amssymb,amsmath}
\usepackage{textcomp}
\usepackage{subfigure}
\usepackage{cite}
\usepackage{braket}
\usepackage{xcolor}
\usepackage{cleveref}
\usepackage{slashed}
\usepackage{stackrel}
\usepackage[0]{editing}

\def\pacs#1{\vspace{10pt} \hspace{0.33cm} \rm PACS numbers: #1 \par \vspace{10pt}}

\title{Bose-Einstein condensation \\ and \\ non-extensive statistics}
\author{ E. Meg\'{\i}as$^{1}$, V. S. Tim\'oteo$^{2}$, A. Gammal$^{3}$ and A. Deppman$^{3}$}
\date{ 
1- Departamento de F\'{\i}sica At\'omica, Molecular y Nuclear and 
Instituto Carlos I de F\'{\i}sica Te\'orica y Computacional, Universidad de Granada 
Avenida de Fuente Nueva s/n, 18071, Granada, Espa\~na \\
2- Grupo de \'Optica e Modelagem Num\'erica, Faculdade de Tecnologia \\
GOMNI/FT - Universidade Estadual de Campinas - UNICAMP \\  
13484-332, Limeira, SP, Brasil \\
3- Instituto de F\'isica, Universidade de S\~ao Paulo \\
05508-090, S\~ao Paulo, SP, Brasil}

\begin{document}

\maketitle

\bigskipamount=1cm

\begin{abstract}
We study the Bose-Einstein condensation in non-extensive statistics for a free gas of bosons, and extend the results to the non-relativistic case as well. We present results for the dependence of the critical temperature and the condensate fraction on the entropic index, $q$, and show that the condensate can exist only for a limited range of $q$ in both relativistic and non-relativistic systems. We provide numerical results for other thermodynamics quantities like the internal energy, specific heat and number fluctuations. We discuss the implications for high energy physics and hadron physics. The results for the non-relativistic case can be of interest in cold-atom systems.
\end{abstract}


\section{Introduction}

One of the most important results of the early Quantum Mechanics is the Bose-Einstein Condensate (BEC). It is a purely quantum effect, however its implications appear at scales similar to those of the classical systems. The description of this phenomenon results from the application of the standard Boltzmann-Gibbs (BG) Statistics to quantum systems~\cite{BEC-trappedgas,Huang}.

Our knowledge on the statistical aspects of mechanical systems has evolved fast in the last decades. The introduction of non-additive entropy extended the reach of the statistical methods to new domains, where systems with complex structure, long-range interaction or following non-Markovian dynamics are found. Several new entropic formulas have been proposed~\cite{Beck-Cohen-Superstatistics}, but the Tsallis's Entropy~\cite{Tsallis,TsallisBook}, $S_q$, is the most ubiquous. The generalized entropy leads to a non-extensive thermodynamics that reduces to the standard Boltzmann-Gibbs theory when the entropic parameter, $q$, approaches the unit~\cite{Tsallis-Curado}. Although the reasons for the large number of systems following the Tsallis statistics are not yet clear, a recent result on the mechanism for the emergence of the non-additive entropy can give some clues on the subject~\cite{Deppman-PRD-2016,DMM_Physics-2020}.

It was shown that a fractal structure in the energy-momentum space, called thermofractal, can be present in any interaction system described in terms of the Yang-Mills fields~\cite{DMM-PRD-2020}. Since this class of systems includes three of the four known interactions, it is clear that thermofractals can be found in many natural phenomena. Since the thermofractals exhibit the non-additive statistics properties, it may explain why systems following the $S_q$ are so commom.

While BEC has been exhaustively studied under the light of the BG Statistics~\cite{Bagnato-Pritchard-Kleppner,Ketterle-vanDruten-PRL-1996,Bagnato-Kleppner}, the same does not hold for the Tsallis Statistics. Apart from a few works on the $S_q$ for quantum systems~\cite{DMM-2015,SukanyaMitra,Rajagopal-Lenzi,Lenzi-Rajagopal,RepulsiveInteractionBEC}, and some applications for BEC, there is a lack of information on the behaviour of the condensates following the non-additive entropy. In this work we provide a detailed description of the non-extensive BEC, or qBEC. We investigate the constraints on the entropic parameter, $q$, for the formation of a condensate, and evaluate the behaviour of the system near the critical temperature.

Although $S_q$ is associated with systems of interacting particles or strongly correlated systems, these aspects do not preclude the formation of BEC~\cite{Penrose-Onsager,Bagnato-Pritchard-Kleppner,Arnaldo2001}, even for systems with finite number of particles~\cite{Bagnato-Kleppner,Ketterle-vanDruten-PRL-1996,Ketterle-vanDruten-PRA-1996}.
In this regard, it is interesting to observe that small systems will follow $S_q$, instead of BG statistics\cite{Lima-Deppman-2020, T-Biro1}, nevertheless, most of the works are restricted to Boltzmann-Gibbs statistics, without considering the possibility of the existence of BEC in the non-extensive  Tsallis Statistics applied to quantum systems.

Despite a relative lack of information on the non-extensive BEC, there are many applications fo the concept in different areas, as in High Energy Physics and Hadron Physics, where the critical temperature of the phase-transition from the confined to the deconfined quark regimes are associated to the formation of a condensate~\cite{BE-FD-interferometry-Particles,Kharzeev-Levin-Tuchin}, or in studies of the Neutron Star structure, where a phase-transition to the deconfined regime might appear~\cite{BEC-HighDensityQuarkGluon}. A detailed study of the constrains for the formation of the condensate and its characteristics is needed. \add{Since one of the motivations for the present work is the possibility of Bose-Einstein Condensate in the high energy collisions, we adopt the relativistic description, which can be straightforwardly restricted to the non-relativistic case, as commented in the present work.}

In section 2 we briefly review the basic concepts associated with BEC in the BG statistics, and in section 3 we review the most important aspects of $S_q$ for the present work. In section 4 we describe the BEC in the non-extensive statistics, and in section 5 we present our conclusions.
In this work, we adopt the natural units, with $c=\hslash=k_B=1$.

\section{Bose-Einstein condensate in Boltzmann-Gibbs statistics} 
\label{BG-BEC}

The thermodynamical potential for a  relativistic gas of massless bosons in the Boltzmann-Gibbs statistics is given by~\cite{DMM-2015}
\begin{equation}
 \ln Z(T,V,\mu)= - \frac{V}{2\pi^2 } \int_0^{\infty} d\varepsilon \, \varepsilon^2 \ln\left[1-\exp[-\beta (\varepsilon-\mu)]\right]\,,
\end{equation}
where $\beta = 1/T$ is the inverse of the temperature. Using the thermodynamical relation
\begin{equation}
 \langle N \rangle = \beta^{-1} \frac{\partial}{\partial \mu} \ln Z \Big|_{\beta} \,,
\end{equation}
we obtain the occupation number of particles,
\begin{equation}
  N(T,V,\mu)= \frac{V}{2\pi^2 }  \int _0^{\infty}  d\varepsilon \, \frac {\varepsilon^{2}}{\exp[\beta (\varepsilon-\mu)]-1}\,. \label{NexcitedNcondensate}
\end{equation}

The occupation number in the equation above diverges at the ground-state energy, $\varepsilon=0$ in the limit $\mu \rightarrow 0$, therefore an infinitesimal vicinity around this value is excluded from the integration and added separately, and we get
\begin{equation}
  N(T,V,\mu) \equiv   N_0 + N_\varepsilon =  \frac{1}{e^{-\beta \mu} - 1}  + \frac{V}{2 \pi^2  }  {\mathcal P} \int _0^{\infty}  d\varepsilon \, \frac {\varepsilon^2}{\exp[\beta (\varepsilon-\mu)]-1} \,, \label{eqn:numberparticlesBG}
\end{equation}
with the first term being the number of particles in the ground-state, $N_0$, which is added separately. The symbol ${\mathcal P}$ in front of the integration sign is used to remember that the singular point is removed, and this integral corresponds to the number of particles in the excited states, $N_{\varepsilon}$. The integration can be computed numerically, and is associated to a family of integrals of the type
\begin{equation}
 g_n(\mu)=\frac{1}{\Gamma(n)}\int _0^{\infty}  d\varepsilon \, \frac {\varepsilon^{n-1}}{\exp[\beta (\varepsilon - \mu)]-1} \, .
\end{equation}
The analysis of the family of function described by the equation above shows that, for $n>1$, 
\begin{equation}
g_n(\mu)=\beta^{-n} \textrm{Li}_n\left( e^{\beta\mu}\right)\,,
\end{equation}
where $\textrm{Li}$ is the polylogarithm function. In the case $\mu=0$, one gets $g_n(0) = \beta^{-n} \zeta(n)$, where $\zeta$ is the zeta function. When the function is limited, the maximum of $g_n(\mu$) happens for $\mu=0$. Thus, $N_{\varepsilon}$ remains finite and~\Cref{eqn:numberparticlesBG} gives the number of particles in the system. The ratio $N_0/N$ is called condensate ratio, and is the most relevant quantity in the present work. When $\mu \rightarrow 0$, the maximum number of particles in the excited states is reached at a critical temperature, $T_c$, and is
\begin{equation}
 N_{\varepsilon}(T)   \le N_{\varepsilon,\textrm{max}}(T_c) \equiv \frac{VT_c^{3}}{\pi^2  }  \zeta(3)\,.
\end{equation}
This is the maximum number of particles allowed in the excites states for the system at temperature $T_c$, and represents the total number of particles if $N_0=0$. The value for $\zeta(3)$ is finite, so the number of particles in the excited state, $N_{\varepsilon}(T)$ remains finite. Below the critical temperature, $T_c$, however, the number of particles allowed in the excited states, $N_{\varepsilon}(T)$,  becomes smaller than the maximum number of particles at the critical temperature, so the excess of particles must be at the ground-state. If we consider that the total number of particles equals the maximum number of particles allowed at the critical temperature, from the equation above, we get 
\begin{equation}
N=N_{\varepsilon,\textrm{max}}(T_c)=\frac{VT_c^{3}}{\pi^2 }\zeta(3)\,,
\end{equation}
therefore
\begin{equation}
T_c=\left(\frac{N}{V}\right)^{1/3}\frac{\pi }{[\pi \zeta(3)]^{1/3}} \,. \label{eq:Tc_BG}
\end{equation}
 The ratio
 \begin{equation}
   \frac{N_0(T)}{N}=\frac{N-N_{\varepsilon}(T)}{N}=1-\left(\frac{T}{T_c}\right)^{3}\,. \label{eqn:relatBG-BEC}
 \end{equation}
Note that we obtain cubic behaviour in the present case because we are considering a relativistic gas, what changes the phase-space topology. For a non-relativistic gas, we would have, instead of~\Cref{eqn:relatBG-BEC}, a similar equation with power $3/2$ in the temperature.

We will study also other quantities like the variance of the condensate population
\begin{equation}
    \Delta N_0^2 = \beta^{-1} \frac{\partial}{\partial \mu} N_0 = \frac{e^{-\beta\mu}}{\left( e^{-\beta\mu} - 1 \right)^2} \,, 
\end{equation}
the total energy of the system
\begin{equation}
    U = - \frac{\partial}{\partial\beta} \ln Z \Big|_{\mu} + \frac{\mu}{\beta} \frac{\partial}{\partial\mu} \ln Z \Big|_{\beta} = \frac{V}{2\pi^2 }  \int _0^{\infty}  d\varepsilon \, \frac {\varepsilon^{3}}{\exp[\beta (\varepsilon-\mu)]-1} = \frac{3V}{\pi^2} g_4(\mu) \,, \label{eq:U} 
\end{equation}
and the specific heat at constant volume
\begin{eqnarray}
    C_V = \frac{\partial U}{\partial T}  \Bigg|_{V} &=&  \frac{V}{2\pi^2} \beta^2  \int _0^{\infty}  d\varepsilon \, \varepsilon^{3} (\varepsilon - \mu)  \frac {\exp[\beta(\varepsilon - \mu)]}{\left( \exp[\beta (\varepsilon-\mu)]-1 \right)^2} \nonumber \\
    &=& \frac{3V\beta}{\pi^2} \left[ 4 g_4(\mu) - \mu g_3(\mu) \right]\,. \label{eq:cv}
\end{eqnarray}

\section{Bose-Einstein condensation in non-extensive statistics}

Tsallis proposed a generalization of the Boltzmann-Gibbs statistics by introducing a new entropy formula, given by
\begin{equation}
 S_q= k  \frac{1-\sum_ip_i^q}{q-1}= -k \sum_i p^q_i \ln_q p_i   \,, \label{eqn:entropy}
\end{equation}
where $p_i$ is the probability, $q$ is a parameter called entropic index, and
\begin{equation}
    \ln_q z = \frac{z^{1-q}-1}{1-q}
\end{equation}
is the q-logarithm function. The new entropic formula has two remarkable facts: it allows to obtain a complete thermodynamic description of the system through the Maximum-Entropy Principle; and it is non-additive.

The inverse of the q-logarithmic function is the q-exponential, that is common in the distributions of systems described by the non-extensive thermodynamics that is derived from the non-additive entropy.

\subsection{Non-extensive thermodynamics} 
\label{scn:non_ext_thermo}

The non-extensive thermodynamics of quantum systems has been studied in~\cite{DMM-2015,SukanyaMitra}. It has been applied to investigate some systems, specially hadronic systems~\cite{Nematollahi}, with large implications in the study of, e.g., can be hadrons, quark-gluon plasma and neutron-stars. It is interesting to notice that the running-constant at the non-perturbative regime has been described by using the non-extensive statistics. The connections between the running-constant and the $S_q$ has been demonstrated in~\cite{DMM-PRD-2020}. Here, BEC in the non-extensive statistics, using as starting point is the non-extensive quantum ideal gas, which was described in reference~\cite{DMM-2015}. 

The non-extensive  entropy is given by the standard thermodynamics relation
\begin{equation}
 S=-\beta^2 \left. \frac{\partial}{\partial \beta}\left(\frac{\ln Z_q}{\beta}\right)\right|_{\mu} \,,
\end{equation}
where $\beta=1/(kT)$, with $k$ being the Boltzmann constant. For the non-extensive ideal gas of bosons or fermions, the thermodynamical potential is given by
\begin{equation}
 \ln Z_q(V,T,\mu)=-\xi V \int \frac{d^3p}{(2\pi)^3} \sum_{r=\pm}\Theta(rx) \ln^{(-r)}_q\left(\frac{e_q^{(r)}(x)-\xi}{e_q^{(r)}(x)}\right)\,. \label{eqn:thermdynamicalpotential}
\end{equation}
In the equations above, we have used the q-exponential functions
\begin{equation}
 \begin{cases}
   & e_q^{(+)}(x)=[1+(q-1) x]^{\frac{1}{q-1}} \,, \qquad   x \ge 0 \\
   & e_q^{(-)}(x)=[1-(q-1) x]^{-\frac{1}{q-1}} \,, \quad\; x < 0 
 \end{cases} \,,
\end{equation}
and the q-logarithm functions, $\ln_q^{(+)}(x)$ and $\ln_q^{(-)}(x)$, defined as the inverse function of, respectively, $e_q^{(+)}(x)$ and $e_q^{(-)}(x)$. The parameter $\xi$ distinguishes the cases of bosons, with $\xi=1$, and fermions, with $\xi=-1$. Here we are interested in the first case, so we fix $\xi=1$. Also, we will deal with positive energies, $\varepsilon \ge 0$, and negative chemical potentials, $\mu \le 0$, so $x := \beta(\varepsilon-\mu) \ge 0$, where $\varepsilon$ is the single-particle state energy. Then, the relevant q-functions will be $\ln_q^{(-)}(x)$ and $e_q^{(+)}(x)$. 

The properties of the q-exponential function have been investigated in many works, and many of those properties are embedded in the so-called q-calculus~\cite{BORGES-qCalculus}. An important property is the duality $q \leftrightarrow q'=2-q$, which is very common to be observed in non-extensive systems. This duality was investigated in  Ref.~\cite{Deppman-Physics-2021} in the context of the $q$-reflection transformation that changes $q=1+\delta$ to $q'=1-\delta$, and it was shown that this transformation is isomorphic to a reflection of the function domain, $x \rightarrow -x$. In these regards, it is interesting to mention that for quantum relativistic systems in the non-extensive statistics, where particles pair creation is allowed, a complete description of the system in thermodynamical terms is possible only with the inclusion of distributions with both $q$ and $q'=2-q$~\cite{Rozynek-Wilk-quasiparticle}.


\subsection{Bose-Einstein condensation and non-extensive effects}

Using the thermodynamical relation
\begin{equation}
 \langle N \rangle = \beta^{-1} \frac{\partial}{\partial \mu} \ln Z_q \Big|_{\mu} \,,
\end{equation}
we obtain the single-particle state occupation number for a bosonic gas,
\begin{equation}
 n_q^{(+)}(\varepsilon,\beta,\mu)=\left[e_q^{(+)}[\beta (\varepsilon-\mu)]-1 \right]^{-q} \,.
\end{equation}

We will investigate in what conditions the singularity corresponding to the Bose-Einstein condensation is present in the formula above. Let $\varepsilon_c$ be the single-particle energy  that is the closest to the chemical potential value, $\mu$. Following the standard approach, we consider this case as separate from the other states, and its occupation number will be
\begin{equation}
 n_q^{0}(\varepsilon_c,\beta,\mu)=\left[e_q^{(+)}[\beta (\varepsilon_c-\mu)]-1 \right]^{-q}\,,
\end{equation}
while the number of particles in the other states are
\begin{equation}
 n_q^{\varepsilon}(\varepsilon,\beta,\mu)=\left[e_q^{(+)}[\beta (\varepsilon-\mu)]-1 \right]^{-q}\,.
\end{equation}
Hereby, the total number of particles is
\begin{equation}
 N_q \equiv N_q^0 + N_q^{\varepsilon} = \left[e_q^{(+)}[\beta (\varepsilon_c-\mu)]-1 \right]^{-q} + \frac{V}{2 \pi^2 }  \int_0^\infty d\varepsilon \, \varepsilon^2 \left[e_q^{(+)}[\beta (\varepsilon-\mu)]-1 \right]^{-q} \,. \label{eqn:Nq}
\end{equation}
If we consider $\mu \rightarrow 0$, the singularity in the occupation number corresponds to the ground-state, $\varepsilon_c=0$.

A careful functional analysis must be carried out in order to verify if the condensate can be formed in the non-extensive statistics, and what are the conditions and the qBEC behaviour in the positive case. For simplicity we introduce $x=\beta(\varepsilon-\mu)$ in the following analysis. The expression for the number of particles in the excited states becomes
\begin{equation}
 N^{\varepsilon}_q = \frac{V}{\pi^2   } \beta^{-3} \, \zeta_q(\mu\beta) \,,
\end{equation}
where~\footnote{The definition of Eq.~(\ref{eq:zeta_q}) corresponds to values $q \ge 1$. The definition of $\zeta_q(\mu\beta)$ for $q < 1$ includes an upper limit in the integral: $\int_{-\mu\beta}^\infty dx \to \int_{-\mu\beta}^{1/(1-q)} dx$.}
\begin{equation}
 \zeta_q(\mu\beta) = \frac{1}{2} \int_{-\mu\beta}^\infty dx \, (x + \mu\beta)^{2} \left[e_q^{(+)}(x)  - 1 \right]^{-q}\,. \label{eq:zeta_q}
\end{equation}
Notice that the maximum number of particles allowed at the critical temperature turns out to be 
\begin{equation}
N_{q,\textrm{max}}^{\varepsilon}(T_c) = \frac{V T_c^3}{\pi^2} \zeta_q(0) \,,
\end{equation}
and therefore the critical temperature is
\begin{equation}
    T_c = \left( \frac{N_q}{V}\right)^{1/3} \frac{\pi}{\left[ \pi \zeta_q(0) \right]^{1/3}} \,. \label{eq:Tc_Tsallis}
\end{equation}

\begin{figure}[t]
 \hspace{4cm} \includegraphics[width=0.47\textwidth]{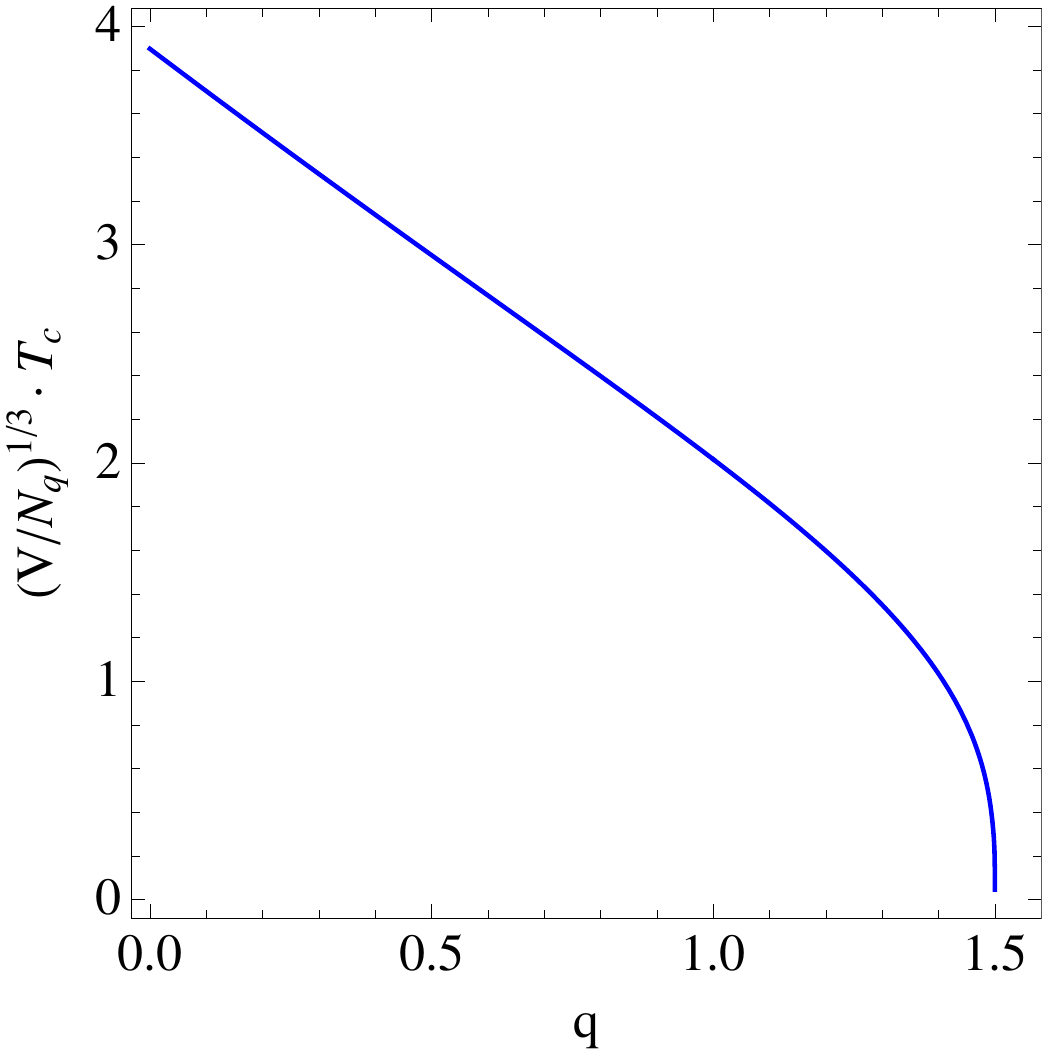} 
 \caption{Critical temperature $(\times (V/N_q)^{1/3})$ as a function of the entropic index $q$. Notice that the curve is independent of the values of $V$ and $N_q$.
 }
 \label{fig:Tcq}
\end{figure}

This expression is the generalization of Eq.~(\ref{eq:Tc_BG}) according the non-extensive statistics. Observe that $\zeta_{q=1}(0) = \zeta(3)$. In~\Cref{fig:Tcq} we show the behaviour of the product $(V/N_q)^{1/3}T_c$, which depends only on the entropic parameter $q$. The product decreases almost linearly up to the vicinity of the critical value $q_c=3/2$. We will show below that this represents the maximum value for the formation of the BEC in non-extensive systems, and that at this limit, the condensate is formed only at temperatures near $T=0$, and the low temperature is a manifestation of a phenomenon that will be investigated in details here, the fact that systems with higher entropic parameters resist to the formation of the condensate, while systems with low values of $q$ are more favourable to the formation of the condensate.

Before studying the physical properties of the qBEC, we will study what are the conditions for it to be formed. The result obtained for the critical temperature already suggests some critical value, and we will show here how it arises in the non-extensive statistics. Considering the q-exponential function in its first order in the Taylor's expansion around the value $x=0$, the integral above for $\mu\beta = 0$ becomes $(\zeta_q \equiv \zeta_q(0))$
\begin{equation}
 \zeta_q= \frac{1}{2} \int_0^\infty dx\, x^{2-q} dx \,, 
 \label{eqn:IRsingularity}
\end{equation}
which remains finite in the limit $x \rightarrow 0$ if $q<3$.  But we have to investigate the limit $x \rightarrow \infty$. In this case we have
\begin{equation}
  \int^{x_+} dx \, x^{2} \left[e_q^{(+)}(x) - 1 \right]^{-q} \rightarrow \frac{(q-1)^{-\frac{1}{q-1}}}{3-2q} x_+^{3-\frac{q}{q-1}}  \qquad (q > 1) \,,  \label{eq:intxp}
\end{equation}
which remains finite in the limit $x_+ \to \infty$ only if $q < 3/2$. Therefore, the number of particles in the excited states will remain finite if $1<q<3/2$. In fact, the case $q=1$ leads also to finite values, as this corresponds to the Boltzmann-Gibbs condensate which was addressed in~\Cref{BG-BEC}. Let us mention that in the case $q < 1$, the integral in the lhs of Eq.~(\ref{eq:intxp}) has the upper limit $x_+ = 1/(1-q)$, and it turns out to be finite for any value of $q < 1$. Then, the range of values of $q$ in which $N_q^\varepsilon$ is finite can in fact be extended to any value $q < 3/2$. These results explain the limiting value for the parameter $q$ observed in~\Cref{fig:Tcq}. Therefore, the qBEC can be obtained only if $q <3/2$.

Observe that $\zeta_q$ is independent of the temperature, so we can calculate the condensate ratio in a way similar to the one used in the BG condensate. We observe that there is a critical temperature for which the number of particles in the condensate, $N_q^0$, is still much smaller than the total number $N_q$, and the number of particles in the excited states is maximum. Below this temperature, any new particle will populate the condensate. In this case,
\begin{equation}
 \frac{N_q^0}{N_q}=1-\frac{N_q^{\varepsilon}}{N_q}=1-\left(\frac{T}{T_c}\right)^{3}\,.
\end{equation}

\begin{figure}[t]
 \begin{subfigure}{}
  \includegraphics[width=0.47\textwidth]{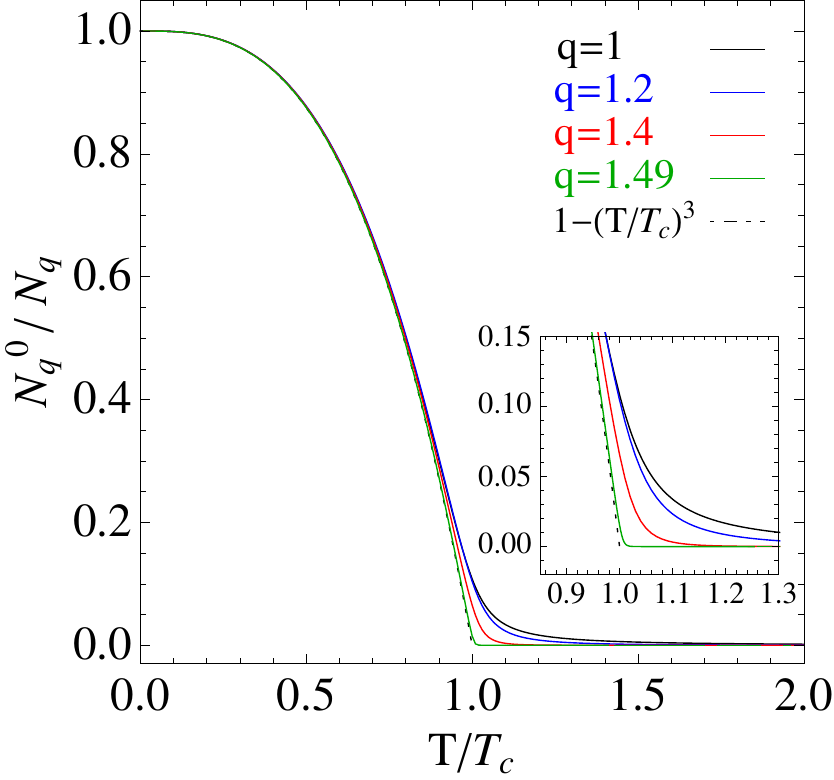} 
 \end{subfigure}
 \begin{subfigure}{}
  \includegraphics[width=0.47\textwidth]{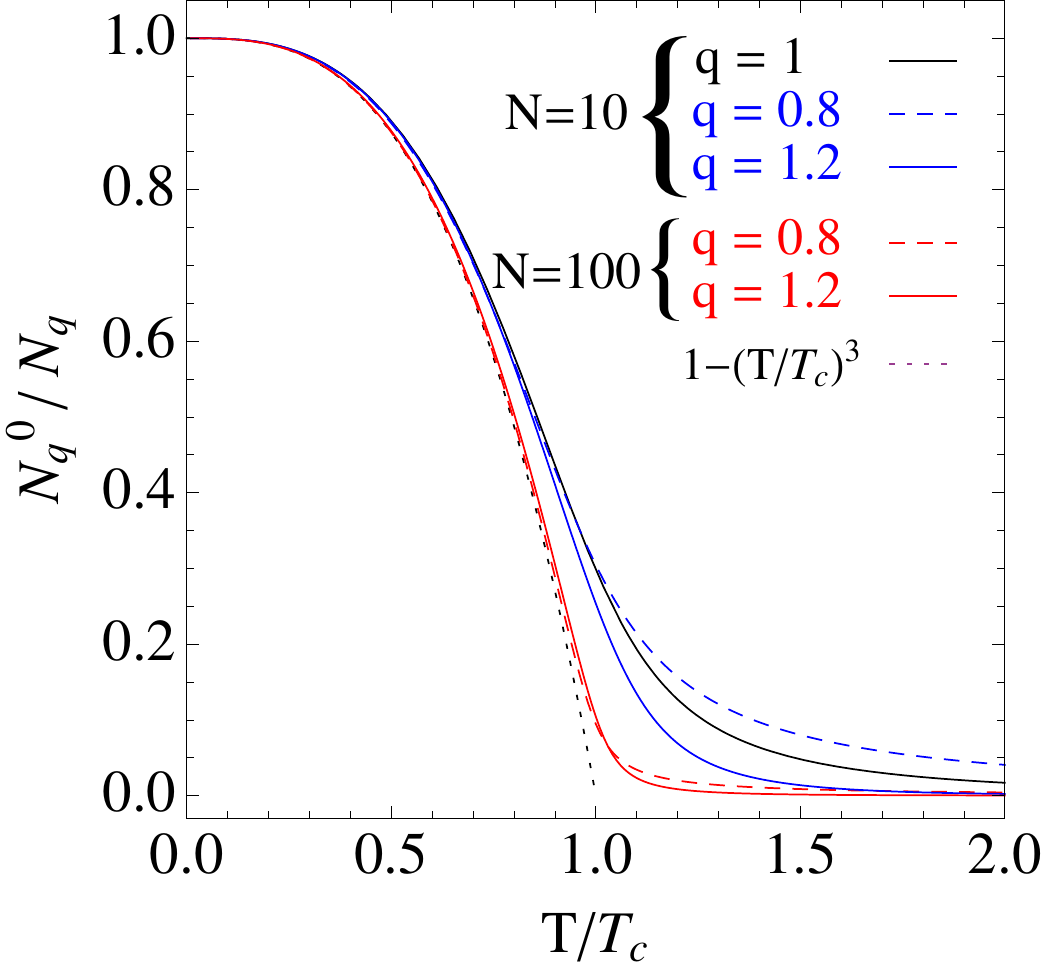} 
 \end{subfigure}
 \caption{Bose-Einstein condensation in the non-extensive statistics. Left panel: plot for fixed value of the number of particles, $N = 100$ and different values of the entropic index, $q$. Right panel: plot for fixed values of the entropic index, $q=0.8$ and $1.2$, and different number of particles.}
 \label{fig:Ro}
\end{figure}

The expression for the number of particles in the excited state is, now, slightly different that in~\Cref{NexcitedNcondensate}, since it gives, in the non-extensive case,
\begin{equation}
 N^{\varepsilon}_q=\frac{V}{2\pi^2 } \int_0^\infty d\varepsilon \, \varepsilon^2  \left[e_q^{(+)}[\beta (\varepsilon-\mu)]-1 \right]^{-q} \,.
\end{equation}

Thus, the non-extensive Bose-Einstein condensate presents the same behaviour with the temperature as the Boltzmann-Gibbs consdensate. This result was verified by numerical calculations (see~\Cref{scn:NumericalCalc}), as shown in~\Cref{fig:Ro}, where we see that the condensate fraction approaches the expected behavior as the number of particles increases, for different values of $q$ in the allowed range.

In the left and right panels of Figure~\ref{fig:Ro} we plot the result of $N_q^0/N_q$ as a function of $T/T_c$ performed by using this numerical procedure. We observe in the left panel that, as $q \rightarrow 3/2$, the  ratio $N_q^0/N_q$ approaches the critical line expected in the Boltzmann-Gibbs Statistics, independently of the number of particles or temperature.  In the right panel of this figure it is observed that the same effect is obtained when increasing the number of particles independently of the value of $q$.

The results shown in Figs.~\ref{fig:Tcq}-\ref{fig:Ro} evidence an interesting behaviour of the qBEC that cannot be observed in the Boltzmann-Gibbs BEC. While in~\Cref{fig:Tcq} we observed the resistance of the system to form the condensate as $q$ increases, which is manifested in the lower critical temperatures in those cases. However, in Fig.~\ref{fig:Ro} we notice a behaviour that, at first sight, seems to contradict the previous conclusion, since they show that the limiting behaviour of the qBEC formation is attained by the systems with higher values of $q$ before those with lower values. This contradiction is not real, though, since in the latter figures the condensate fraction is plotted as a function of the ratio $T/T_c$.

The conclusions we can draw from these initial results are that, while the qBEC resistance for the condensate formation increases with $q$, the phase-transition to the condensate is sharper for systems with large values of $q$ than for those system with lower values of $q$. To investigate this behaviour in more details, we will study the dependence of the number of particles in the first excited state on the temperature.

We will show that the analysis of the fraction of particles in the first excited states confirms the conclusions obtained above. A few modifications in our method are necessary, so let us describe them before turning our attention to the physical aspects of the problem. While in~\Cref{eqn:Nq} {the system is supposed to have a continuum of states, one can obtain a discretization of the energy levels when considering it inside a large cubical box of length $L$. Then, the energy levels of relativistic massless particles is
\begin{equation}
E_{n_x, n_y, n_z} = \frac{\pi}{L}  \sqrt{n_x^2 + n_y^2 + n_z^2}  \,, \qquad n_x, n_y, n_z \ge 1  \,. \label{eq:Ebox}
\end{equation}
The results obtained with the method described above are plotted in~\Cref{fig:R1}. We see that there is a peak in the ratio $N^1_q/N_q$ as a function of the relative temperature, $T/T_c$. The presence of the peak is observed for all values of $q$ below $q_c$, and at a position near the critical temperature. As the number of particles increases, the peak becomes narrower. A clear dependence on the value of $q$ is observed, and as $q$ increases the relative number of particles in the first excited state tends to decrease. The shape of the curves are similar in all cases, including for $q=1$ that corresponds to the BG condensate.

The reduction of the number of particles for higher values of $q$ is associated with the sharper phase-transition discussed above, and results from the fact that a larger fraction of the particles are in the ground-sate. This aspect is already present in the Fig.~\ref{fig:Ro}, where the smaller tails at the right side of the critival temperature indicates the predominance of the condensate for those systems with large $q$. Therefore, the results of the analysis of the first excited state confirms the conclusions obtained above.

Another interesting aspect of the system that can be observed only in the non-extensive condensate can be observed in~\Cref{fig:R1}. Comparing the left and right panels of the figure, we note that the distributions with $q=1.0$, $q=1.1$ and $q=1.2$, that are clearly separated in the case of $N=10$, becomes very similar for $N=100$. This result indicates that the relative contributions of the particles in the excited states changes with the number of particles. We will investigate further this aspect of the qBEC.

\begin{figure}[t]
 \begin{subfigure}{}
  \includegraphics[width=0.4\textwidth]{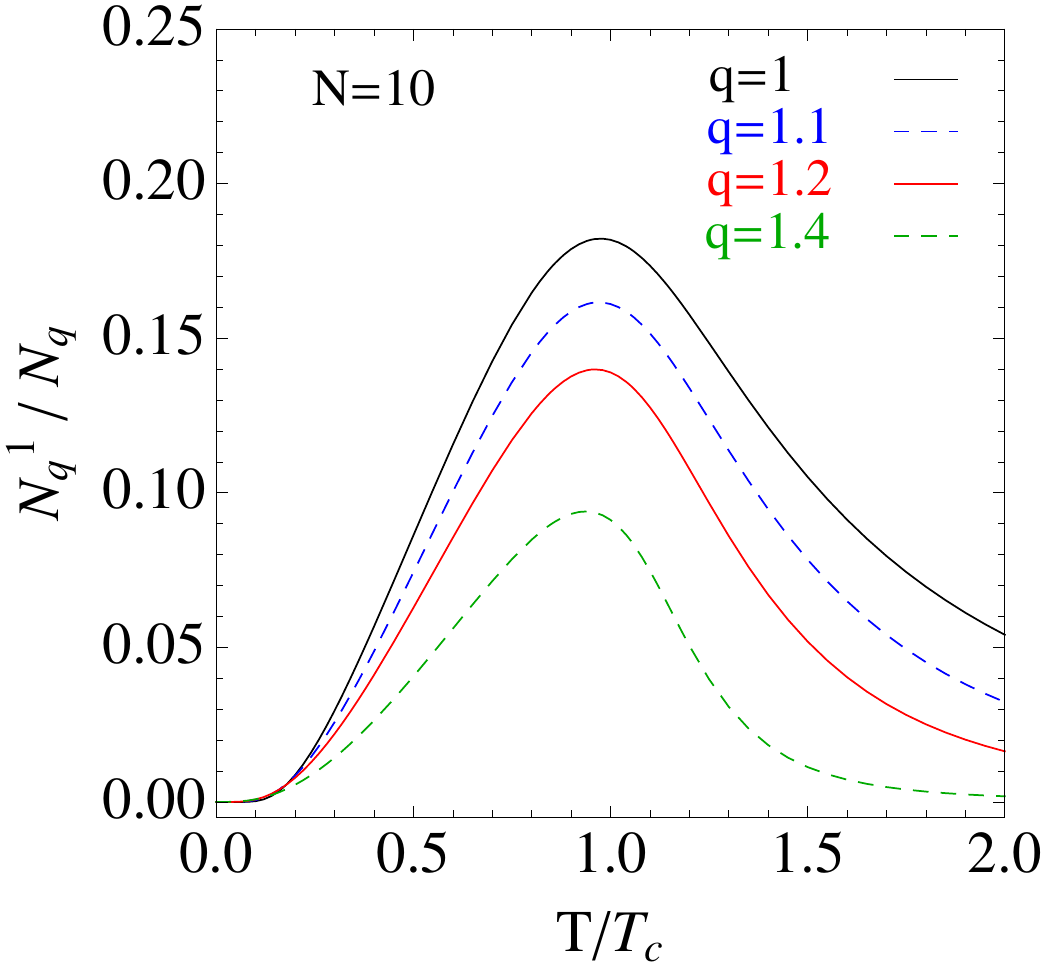} 
 \end{subfigure}
 \begin{subfigure}{}
\hspace{1cm}  \includegraphics[width=0.4\textwidth]{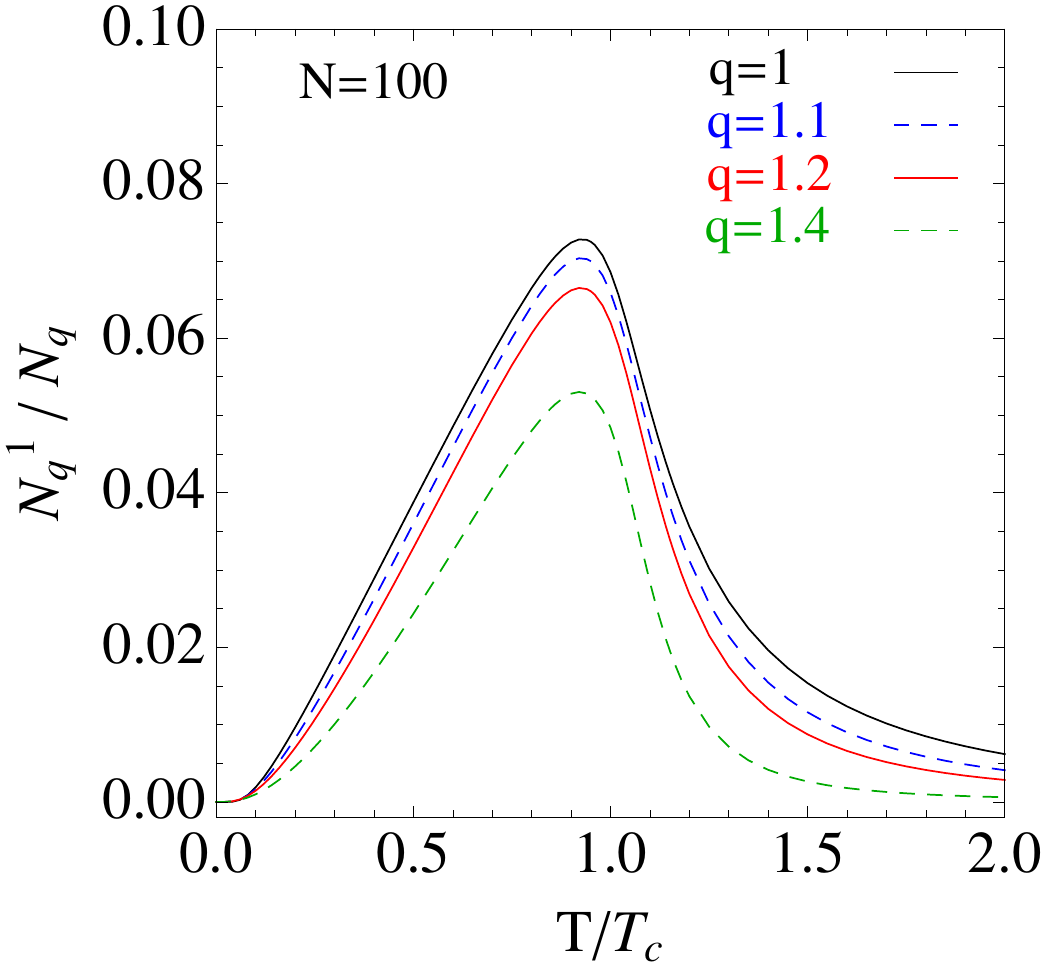} 
 \end{subfigure}
 \caption{Fraction of particles in the first excited state within the non-extensive statistics for fixed values of the number of particles, $N = 10$ (left panel) and $N = 100$ (right panel), and different values of the entropic index, $q$.
 }
 \label{fig:R1}
\end{figure}

\begin{figure}[t]
  \begin{subfigure}{}
  \includegraphics[width=0.41\textwidth]{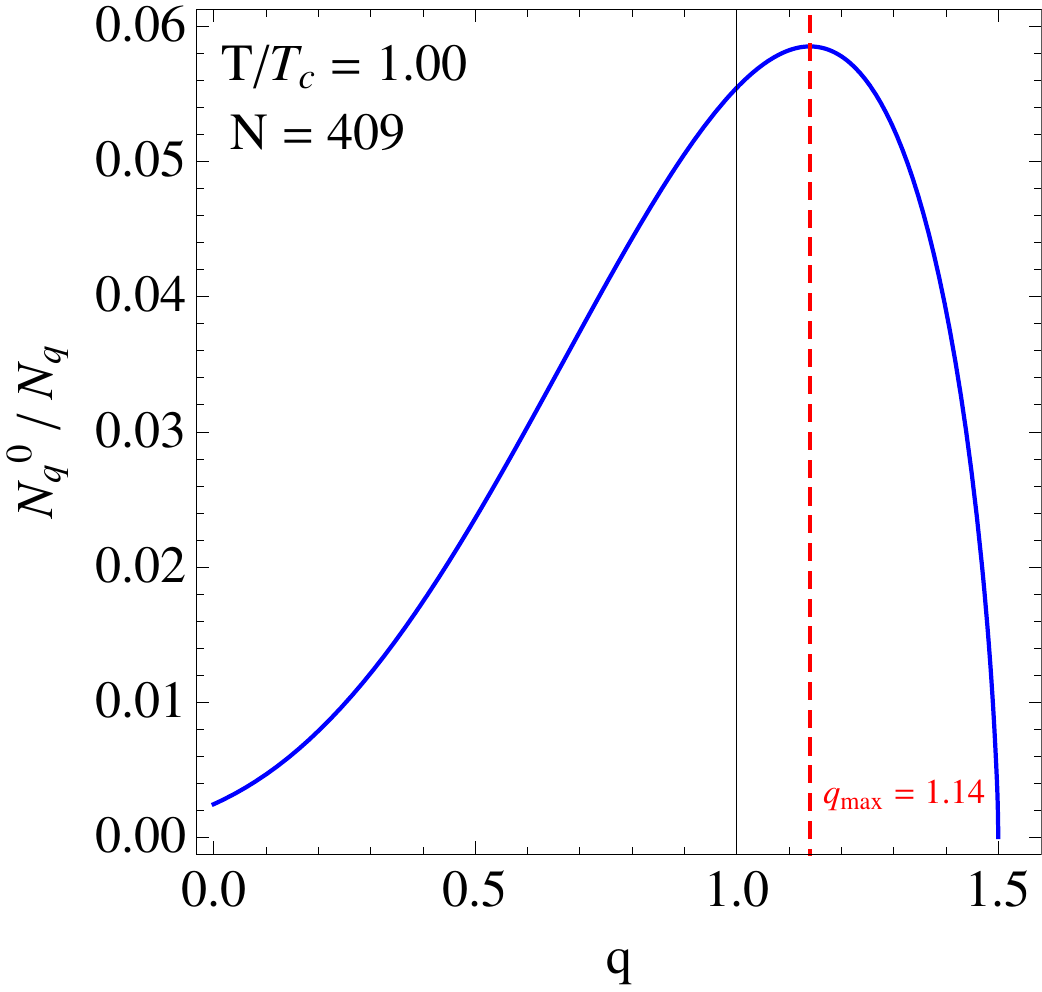} 
 \end{subfigure}
   \begin{subfigure}{}
\hspace{1cm}  \includegraphics[width=0.41\textwidth]{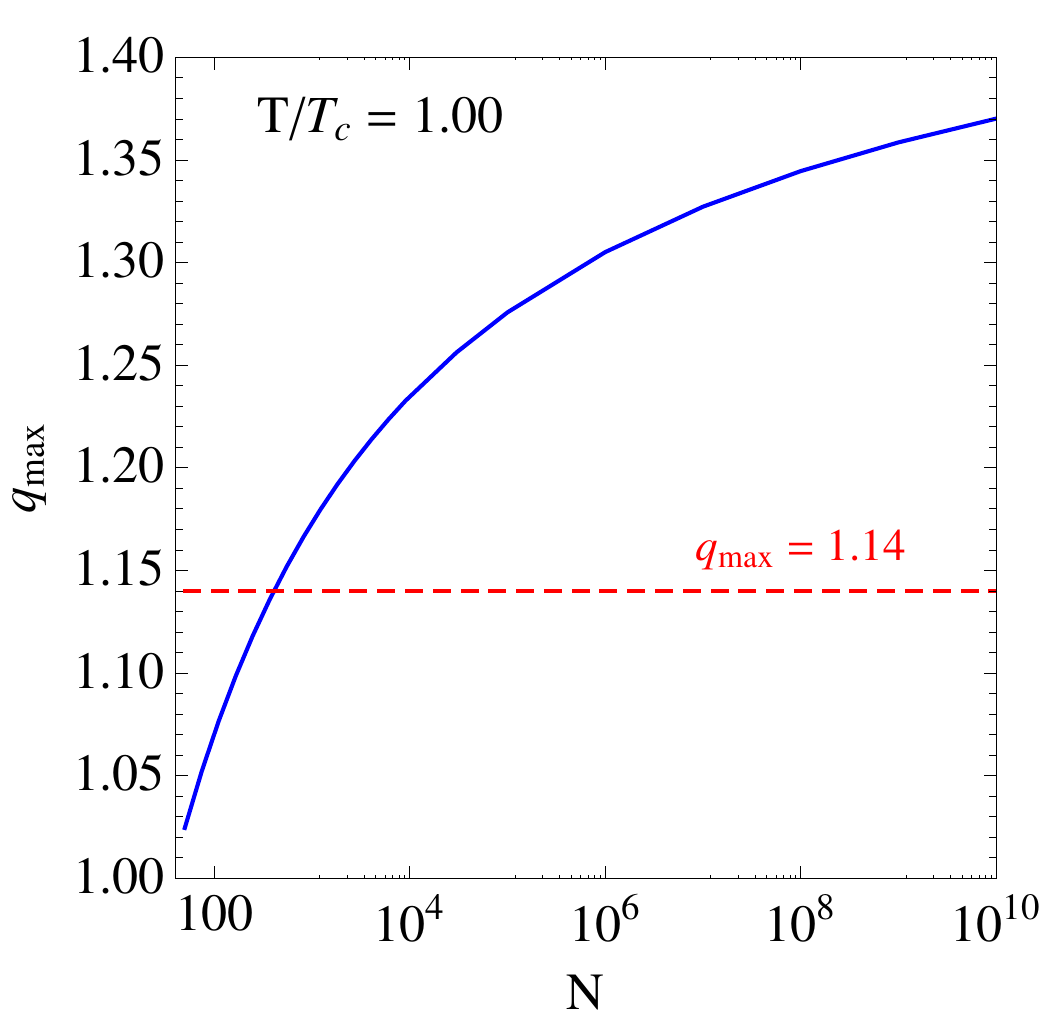} 
 \end{subfigure}
 \caption{Left panel: fraction of particles in the condensate within the non-extensive statistics as a function of $q$. The maximum value of $N_q^0/N_q$ is obtained at $q_{\textrm{max}}$. Right panel: behavior of $q_{\textrm{max}}$ as a function of $N_ q$. The value $q_{\textrm{max}} = 1.14$ is obtained for $N = 409$. It is shown as dashed (red) lines in both panels the curve $q_{\textrm{max}} = 1.14$. We have considered $T/T_c = 1$ in both panels.}
 \label{fig:N0Nq_ratio1}
\end{figure}

The behaviour of the part of the system at the excited states near the critical temperature is complex, therefore we will use some approximations to evaluate how the number of particles in the excited states vary with $0<q<3/2$.  As discussed above, we can divide the integration in two regions, small $x$ and large $x$, and the approximations
\begin{equation}
 \zeta_q \simeq \zeta_{q \, 1}(x_1) + \zeta_{q\, 2}(x_2) 
\end{equation}
with
\begin{equation}
  \begin{cases}
    \zeta_{q\, 1}(x_1) = \frac{1}{2} \int_0^{x_1} x^{2-q} dx \\
    \zeta_{q\, 2}(x_2) = \frac{1}{2} (q-1)^{-\frac{q}{q-1}} \int_{x_2}^{\infty} x^{2-\frac{q}{q-1}} dx
  \end{cases} \label{eqn:approximation}
\end{equation}
are valid for appropriately chosen $x_1$ and $x_2$. We will assume that there is a single value $x_1=x_2=X$ for which the approximations above can be considered valid, where $X$ is to be determined. In the range of values for $q$ mentioned above the integrations can be easily performed, and we get
\begin{equation}
  \begin{cases}
   \zeta_{q\, 1}(X) = \frac{1}{2} \frac{1}{3-q} X^{3-q}  \\
    \zeta_{q\, 2}(X) = \frac{1}{2} \frac{(q-1)^{-\frac{1}{q-1}}}{3-2q} X^{2-\frac{1}{q-1}}
  \end{cases} \,.
\end{equation}
We choose $X$ such that the integrands of the functions defined in~\Cref{eqn:approximation} are continuous. This condition is satisfied only  if 
\begin{equation}
X = (q-1)^{-\frac{1}{2-q}} \,. \label{eq:X}
\end{equation}
With the approximations made above, we obtain the $\zeta_q$ function
\begin{equation}
 \zeta_q= \frac{q(2-q)}{2(3-q)(3-2q)} (q-1)^{-1-\frac{1}{2-q}}\,.
\end{equation}
The number of particles in the excited states is, therefore, given by
\begin{equation}
  N_q^{\varepsilon} = \frac{V}{\pi^2  } \beta^{-3} \frac{q(2-q)}{2(3-q)(3-2q)} (q-1)^{-1-\frac{1}{2-q}} 
  \,.
\end{equation}
The approximation used to obtain this analytical result works better for values of $q$ close to $3/2$, while for $q \lesssim 1.3$ there are some deviations. The equation above indicates that the number of particles would diverge when $q$ tends to $q=3/2$. 

The critical value, $q_c=3/2$, appear also in the analysis of the condensate fraction as a function of $q$. In~\Cref{fig:N0Nq_ratio1}~(left) we observe a peak in the curve at a position $q_{\textrm{max}}$. which depends on the number of particles in the system. The behaviour of $q_{\textrm{max}}$ with $N$ is shown in~\Cref{fig:N0Nq_ratio1}~(right), and we note its continuous increase with the number of particles, asymptotically approaching the value $q_c=3/2$. This plot, indeed, is related to that in~\Cref{fig:Tcq}.

The critical value, $q_c$, is associated with the system for which the condensate can be formed just at $T=0$. On the other hand, the $N$ particle system with $q=q_{\textrm{max}}(N)$ is the one with the sharpest transition to the condensate regime, as evidenced by the maximum ration $N^{0}_q/N_q$.

\begin{figure}[t]
 \begin{subfigure}{}
  \includegraphics[width=0.47\textwidth]{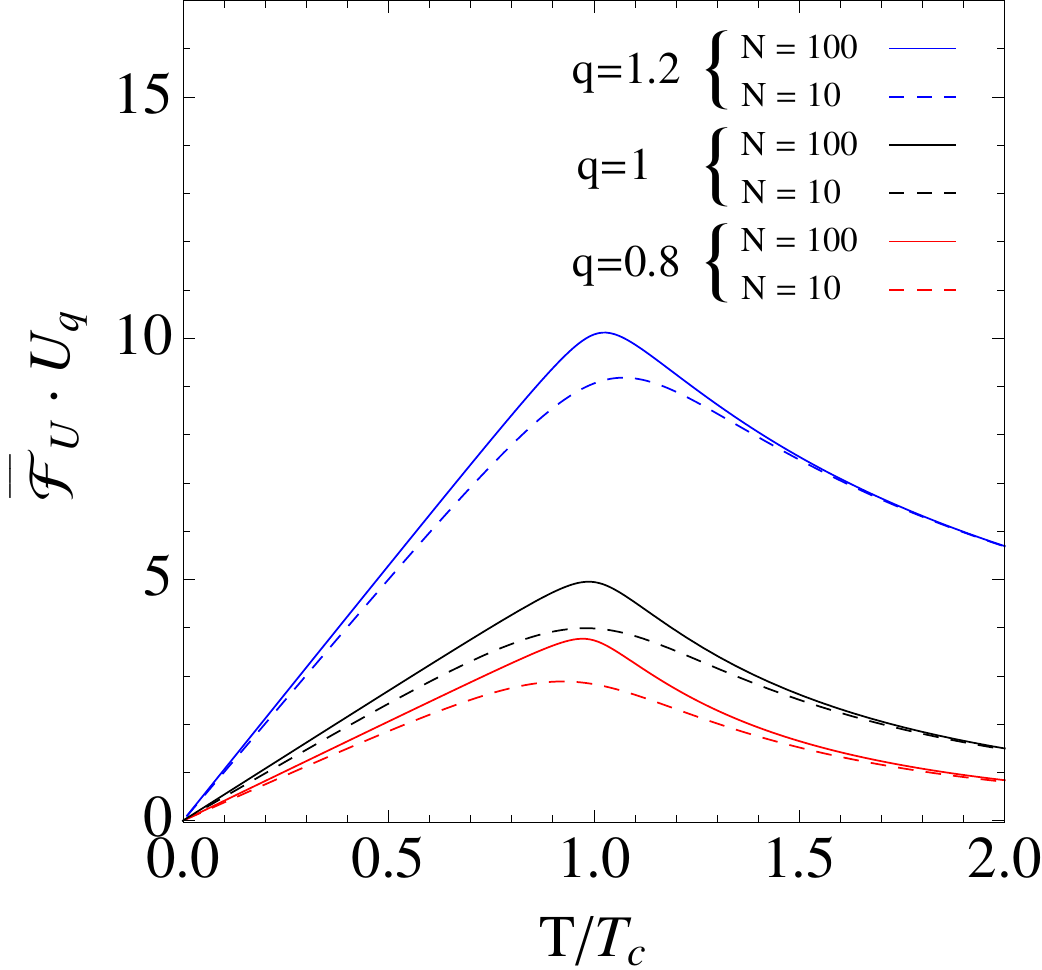} 
 \end{subfigure}
 \begin{subfigure}{}
  \includegraphics[width=0.47\textwidth]{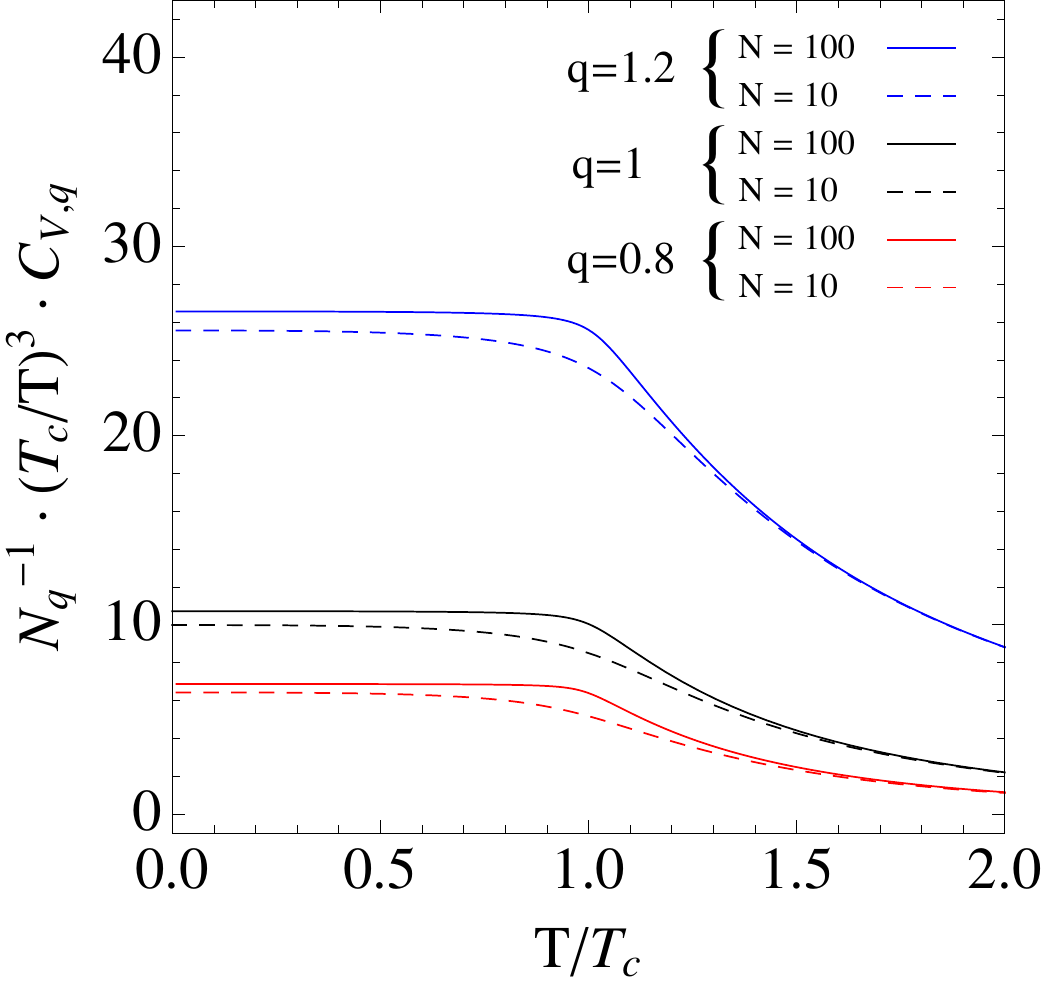} 
 \end{subfigure}
 \caption{Total energy, normalized by the factor $\mathcal {\bar F}_U := N_q^{-1} (V/N_q)^{1/3} (T_c/T)^3$ (left panel), and specific heat at constant volume $(\times N_q^{-1} (T_c/T)^3)$ (right panel), in the non-extensive statistics. We display the results for fixed values of the entropic index, $q = 0.8$ and $1.2$, and different number of particles, $N=10$ and $100$. It is displayed also the results in BG statistics in both panels.
 }
 \label{fig:Uq}
\end{figure}

\subsection{Total energy, specific heat and number fluctuations}

The aspects of the qBEC analyzed so far do not consider the energy of the system. It turns out that the energy imposes some additional constraints on the non-extensive system. 
The total energy of the system is obtained by the standard relation
\begin{equation}
    U_q = - \frac{\partial}{\partial\beta} \ln Z_q \Big|_{\mu} + \frac{\mu}{\beta} \frac{\partial}{\partial\mu} \ln Z_q \Big|_{\beta} 
\end{equation}
resulting in
\begin{equation}
 U_q =\frac{V}{2\pi^2 } \int_0^\infty d\varepsilon \, \varepsilon^3  \left[e_q^{(+)}[\beta (\varepsilon-\mu)]-1 \right]^{-q} \,.
\end{equation}
By using a matching procedure similar to the one presented for $N_q^\varepsilon$, the total energy can be approximated by 
\begin{eqnarray}
 U_q &=& \frac{V\beta^{-4}}{2\pi^2 } \int_0^{\infty} dx \, x^3  \left[e_q^{(+)}(x)-1 \right]^{-q} \nonumber \\
 &=& \frac{V\beta^{-4}}{2\pi^2 } \left[ \int_0^{X} dx \, x^{3-q}  +  (q-1)^{-\frac{q}{q-1}} \int_{X}^{\infty} dx  \, x^{3-\frac{q}{q-1}} \right] \,,  \label{eq:Uq}
\end{eqnarray}
for $X$ given by Eq.~(\ref{eq:X}). The first integral in the last line of Eq.~(\ref{eq:Uq}) converges to a finite value for $q < 4$, but the second integral converges only if $q < 4/3$. The result for $U_q$ is 
\begin{equation}
 U_{q} = \frac{V\beta^{-4}}{2\pi^2 } \frac{q(2-q)}{(4-q)(4-3q)} (q-1)^{-1-\frac{2}{2-q}} \,,
\end{equation}
which turns out to be a good approximation for $1.3 \lesssim q < 4/3$. Therefore, in the limit for $q$ found in the analysis of the condensate fraction, those systems with $q>4/3$ present an infinite energy, despite the finite number of particles. The conclusion is that they do not represent a physical system. 


The energy of the system for $q=0.8$ and $q=1.2$, as a function of $T/T_c$, is shown in the left panel of~\Cref{fig:Uq}. We observe a linear behaviour for temperatures below $T_c$, which is a consequence of the formation of the condensate since the number of particles in the ground-state is proportional to $T$. The energy of the system with $q=0.8$ remains below that for $q=1$, while for $q=1.2$ the energy remains sistematically above the Boltzmann-Gibbs cases.

The system with large $q$ presents a more pronounced change at the critical temperature than those with small $q$. This aspect can be appreaciated also by observing the behaviour of the specific heat at constant volume, $C_{V, q}$  constant volume, that is calculated by
\begin{equation}
    C_{V, q} = \frac{\partial U_q}{\partial T}  \Bigg|_{V} =  \frac{V}{2\pi^2} \beta^2  q \int _0^{\infty}  d\varepsilon \, \varepsilon^{3} (\varepsilon - \mu)  \frac {e_q^{(+)}[\beta(\varepsilon - \mu)]^{2-q}}{\left( e_q^{(+)}[\beta (\varepsilon-\mu)]-1 \right)^{q+1}} \,. \label{eq:CVq}
\end{equation}

In the right panel of~\Cref{fig:Uq} we see that the product $T^{-3} C_{V, q}$ is approximately constant up to the critical temperature and decreases above $T_c$, reflecting the modification in the system structure. There is a clear dependence on $q$ in both $U_q$ and $C_{V, q}$.

Finally, we calculate the variance of the condensate population in the non-extensive statistics for fixed total number of particles, given by
\begin{equation}
    \Delta N_q^{0 \,2} = \beta^{-1} \frac{\partial}{\partial \mu} N_q^0 = q \frac{e_q^{(+)}[-\beta\mu]^{2-q}}{\left( e_q^{(+)}[-\beta\mu] - 1 \right)^{q+1}} \,.
\end{equation}
This is plotted in Fig.~\ref{fig:DeltaNN} for fixed values of the entropic index. Notice that the variance tends to decrease with the value of $q$.  This is an interesting quantitiy, since it can be measured experimentally~\cite{BEC-Variance}.

\add{We note that the non-extensive non-relativistic case can be treated in a similar way, but a few results will be different. The behaviour of the condensate fraction, as given by~\Cref{eqn:relatBG-BEC} will depend on $(T/T_c)^{3/2}$. The range of value for the entropic parameter, $q$, where the BEC can be obtained will be $0<q<3$, while the energy of the system will be finite if $0<q<5/3$. These differences between the relativistic and non-relativistic are due to the topology of the phase-space.
}

\begin{figure}[t]
  \begin{subfigure}{}
  \includegraphics[width=0.47\textwidth]{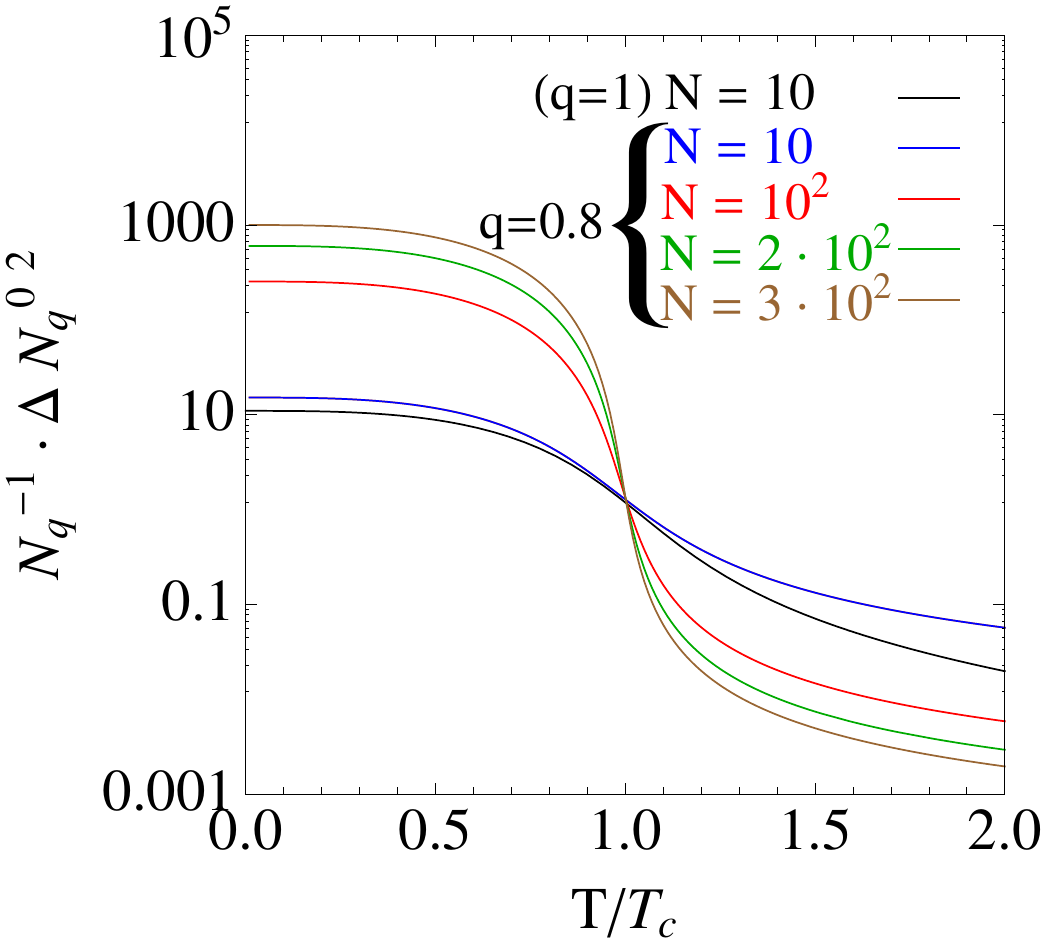} 
 \end{subfigure}
 \begin{subfigure}{}
  \includegraphics[width=0.45\textwidth]{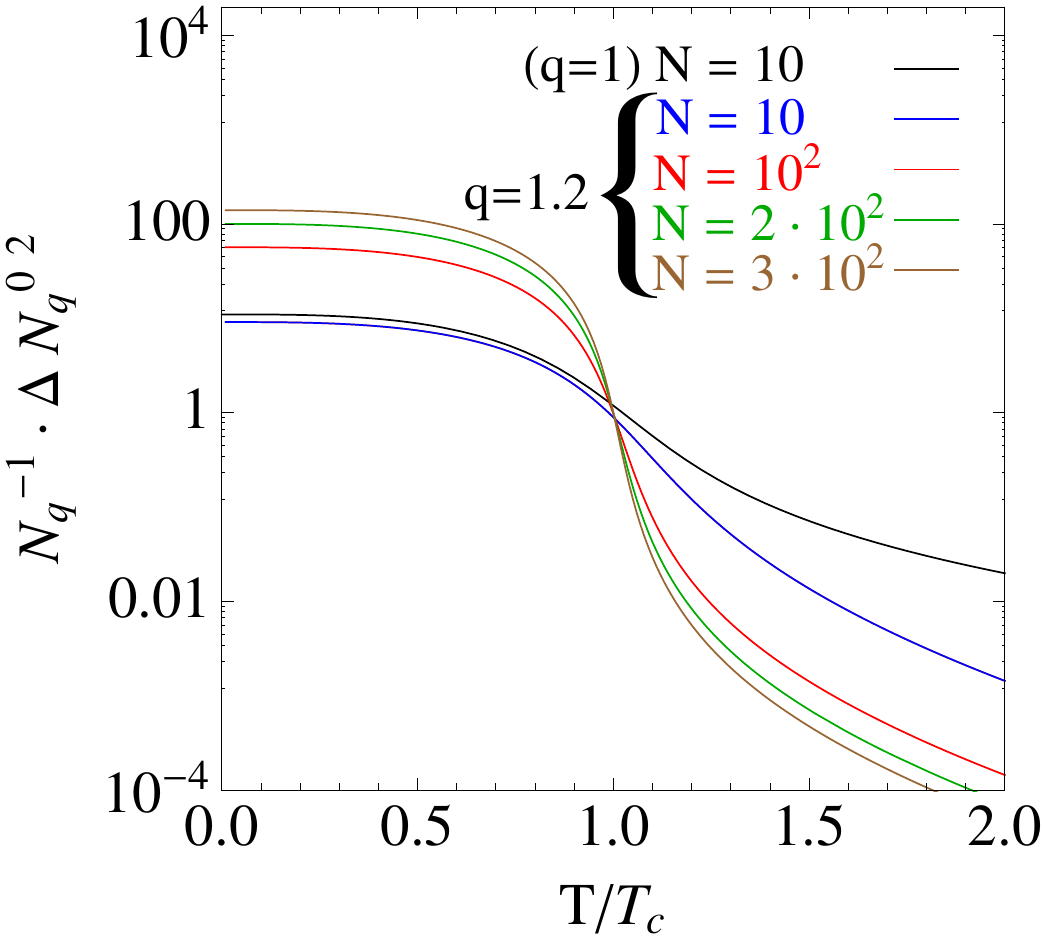} 
 \end{subfigure}
 \caption{Variance of the condensate population in the non-extensive statistics for fixed values of the entropic index, $q=0.8$ (left panel) and $q=1.2$ (right panel), and different number of particles. 
 }
 \label{fig:DeltaNN}
\end{figure}


%

\subsection{Numerical strategy} \label{scn:NumericalCalc}

Apart from these analytical considerations, we have performed a numerical study of Bose-Einstein condensation. The numerical calculations are performed with the following steps:
\begin{enumerate}
 

\item Given a value $\bar N_q$ for the total number of particles, the critical temperature $T_c$ is then computed as the value of temperature such that $\bar N_q = N_{q,\textrm{max}}^{\varepsilon}(T_c)$. This leads to a dependence $T_c \propto (\bar N_q/V)^{1/3}$, a property that was already obtained both in BG and in Tsallis statistics, cf. Eqs.~(\ref{eq:Tc_BG}) and (\ref{eq:Tc_Tsallis}).

\item For the same value of $\bar N_q$ and for a given temperature $T$, usually different from $T_c$, it is computed the value of the chemical potential~$\mu$ such that $\bar N_q = N_q(T,\mu) = N_q^0(T,\mu) + N_q^{\varepsilon}(T,\mu)$.

\item We evaluate the ratios $N_q^0(T,\mu)/\bar N_q$ and $T/T_c$ at temperature $T$. 
\end{enumerate}


Let us mention that the curves of $N_q^0 / N_q$ and $N_q^1/N_q$ as a function of $T/T_c$ are independent on the value of $V$, and the same can be said for the variance $\Delta N_q^{0\, 2}$, the specific heat $C_{V,q}$ and the product $V^{1/3} U_q$.  

\section{Conclusions and outlook}

We studied the Bose-Einstein Condensate in the non-extensive statistics (qBEC), contributing to complement the lack of information on those systems. We calculated the critical temperature and the condensate fraction. These quantities present a dependence on the entropic parameter, $q$. We show that the qBEC can exist in non-extensive system only for $ q < q_c$ with $q_c = 3/2$.

At the critical value, $q_c$, the critical temperature is null. As $q$ decreases from its maximum value, the critical temperature shows, initially, a fast increase, and then, around $q=4/3$, it increases linearly as $q$ decreases. This behaviour was interpreted as a resistance of the system to the formation of the condensate. The large is the entropic parameter, the stronger is the resistance.

On the other hand, those systems with larger values for $q$ present a sharper phase-transition. This was evidenced by the behaviour of the condensate fraction at the critical temperature. The dependence of the condensate fraction with the entropic parameter was investigated in details, and we study it for all values of $0<q<3/2$.

We studied the energy and the specific heat of the qBEC, and showed that the energy presents a divergence for $q>4/3$. Therefore, physical qBEC can exist only in the range $0<q<4/3$. In the condensate regime, the specific heat is constant for all values of $q$, but its value is q-dependent. This dependence arises from the fact that the condensate fraction depends on the value $q$.

We analyzed the variance of the number of particles.

The possibility of formation of a condesnate phase in the hadronic systems have been postulated in several works~\cite{M-Huang-2021,C-Pajares-2021,Kharzeev-Levin-Tuchin,BEC-pion}. Investigations on the High Energy Physics distributions show that, for hadronic systems, $q \sim 1.14$, in good agreement with the theoretical prediction~\cite{DMM-PRD-2020, DMM_Physics-2020}. The results obtained here shows that the QCD allows for the formation of a condesate regime of the gluonic field. The implications of this results go beyond the high energy collisions, reaching the hadron~\cite{BE-FD-interferometry-Particles,Ruggieri-BEC,Andrade-DMM-Nunes-2020} and neutron star structures~\cite{EoS-NS2021,BEC-HighDensityQuarkGluon,Chavanis-BEC-NS,GotsmanLevinMaor-BEC,DMM-Castro-2015}. It is remarkable that the results on the fractal structure of Yang-Mills fields include systems governed by electrodynamic interaction, for which one expects $q<1$, so effects of the qBEC can be investigated in cold atom condensate as well.  The observation of power-law distributions and chaotic behaviour in Bose-Einstein condensate with vertex~\cite{BECvortices-PowerLaw,Arnaldo2,Arnaldo3} can be indicative of non-extensive behaviour in those systems. In these regards, we commented about the extension of the results obtained here to the non-relativistic case.

The transition behaviour observed in the Bose-Einstein condensates when the temperature changes
is of the same nature of the behaviour of the quark condensates when the coupling changes (in a model with only contact interactions). But there is a difference though: quark models are handled
with field theory techniques. We plan to perform a similar study for hadronic condensates trying 
to connect the deviations from the extensive statistics, measured by $q$, to some sort of effective
interaction added to the system.

\vspace{0.5cm}

\section*{Acknowledgements}
The work of E M is supported by the Spanish MINECO under Grant FIS2017-85053-C2-1-P, by the FEDER/Junta de Andaluc\'{\i}a-Consejer\'{\i}a de Econom\'{\i}a y Conocimiento 2014-2020 Operational Program under Grant A-FQM-178-UGR18, by Junta de Andalucía under Grant FQM-225, and by the Consejería de Conocimiento, Investigación y Universidad of the Junta de Andalucía and European Regional Development Fund (ERDF) under Grant SOMM17/6105/UGR. The research of E M is also supported by the Ramón y Cajal Program of the Spanish MINECO 
under Grant RYC-2016-20678. V S T is supported by FAEPEX (grant 3258/19), FAPESP (grant 2019/010889-1) and CNPq 
(grant 306615/2018-5). A D is partially supported by the Conselho Nacional de Desenvolvimento Científico e Tecnológico (CNPq-Brazil), grant 304244/2018-0, by Project INCT-FNA Proc. No. 464 898/2014-5,. A G is supported by CNPq grant PQ 306920/2018-2. A D and A G are  supported by FAPESP grant 2016/17612-7. 


\linespread{.9}
\bibliographystyle{unsrt}
\bibliography{
  Bibliography_TsallisBEC,
  Bibliography_Tsallis,
  Bibliography_Plasma, 
  Bibliography_TsallisHEP, 
  Bibliography_FractalHEP, 
  Bibliography_OurPapers_Thermofractals,
  Bibliography_SelfConsistentHadron,
  Bibliography_Fractals,
  Bibliography_RenormalizationYMF,
  Bibliography_HadronResonanceGasModel,
  Bibliography_ScalingHEP,
  Bibliography_FractalNets,
  Bibliography_HadronBEC
 }

\end{document}